\documentclass{iopart} 
\usepackage{amsmath}
\usepackage[T1]{fontenc}
\usepackage[latin9]{inputenc}
\usepackage{booktabs}
\usepackage{amssymb}
\usepackage{graphicx}
\usepackage{amsfonts}
\usepackage{dcolumn}
\usepackage{bm}
\usepackage{float} 

\begin{document}

\review{Towards a random laser with cold atoms}


\author{W.~Guerin$^1$, N.~Mercadier$^1$,  F.~Michaud$^1$\footnote{Present address: LNE-SYRTE, CNRS UMR 8630, UPMC, Observatoire de Paris, 61 rue de l'Observatoire, 75014 Paris, France},
D.~Brivio$^1$\footnote{Present address: Dipartimento di Fisica,
Universit\`a di Milano, I-20113, Italy}, L.~S.~Froufe-P\'erez$^2$,
R.~Carminati$^3$, V.~Eremeev$^4$, A.~Goetschy$^4$,
S.~E.~Skipetrov$^4$, R.~Kaiser$^1$}


\address{$^1$ Institut Non Lin\'{e}aire de Nice, CNRS and Universit\'{e} de
Nice Sophia-Antipolis,\\ 1361 route des Lucioles, 06560 Valbonne,
France}
\address{$^2$ Instituto de Ciencia de Materiales de Madrid, CSIC, Sor Juana In\'{e}s de la
Cruz~3, Cantoblanco, Madrid 28049, Spain}
\address{$^3$ Institut Langevin, ESPCI ParisTech, CNRS, Laboratoire
d'Optique Physique,\\ 10 rue Vauquelin, 75231 Paris Cedex 05,
France}
\address{$^4$ Universit\'e Joseph Fourier, Laboratoire de Physique
et Mod\'elisation des Milieux Condens\'es, CNRS, 25 rue des
Martyrs, 38042 Grenoble, France}

\ead{William.Guerin@inln.cnrs.fr}


\pacs{37.30.+i,42.25.Dd,42.55.Zz}

\submitto{\JOA}  


\begin{abstract}  
Atoms can scatter light and they can also amplify it by stimulated
emission. From this simple starting point, we examine the
possibility of realizing a random laser in a cloud of laser-cooled
atoms. The answer is not obvious as both processes (elastic
scattering and stimulated emission) seem to exclude one another:
pumping atoms to make them behave as an amplifier reduces
drastically their scattering cross-section. However, we show that
even the simplest atom model allows the efficient combination of
gain and scattering. Moreover, supplementary degrees of freedom
that atoms offer allow the use of several gain mechanisms,
depending on the pumping scheme. We thus first study these
different gain mechanisms and show experimentally that they can
induce (standard) lasing. We then present how the constraint of
combining scattering and gain can be quantified, which leads to an
evaluation of the random laser threshold. The results are
promising and we draw some prospects for a practical realization
of a random laser with cold atoms.
\end{abstract}

\noindent{\it Keywords\/}: Random laser, cold atoms

\maketitle


\section{Introduction: atoms as scatterers and amplifiers}

Among atom-light interactions, elastic -- or Rayleigh --
scattering is one of the simplest processes
\cite{Cohen:PetitLivreRouge-EN}. It has been known for a long time
that in optically thick atomic vapors, multiple scattering leads
to diffusion of light, or ``radiation trapping''
\cite{Holstein:1947,Molisch}. However, only the progress in laser
cooling and trapping \cite{Metcalf} allowed the more recent
demonstration and study of this effect with only true elastic
scattering, by suppressing the Doppler-induced frequency
redistribution \cite{Fioretti:1998,Labeyrie:2003,Labeyrie:2005}.
Many experiments have also been performed to study the coherence
properties of multiply-scattered light in such a medium,
especially using coherent backscattering as a probe
\cite{Labeyrie:1999,Bidel:2002} (see
\cite{Kaiser:2005,Kupriyanov:2006,Labeyrie:2008,Aegerter:2009} for
review articles devoted to this topic).

Cold atoms have also driven a renewed interest in non linear
spectroscopy, as the suppression of Doppler broadening allows the
resolution of narrow spectral resonances in pump-probe
spectroscopic schemes \cite{Grynberg:2001}. One example is Raman
gain between Zeeman sublevels \cite{Tabosa:1991,Grison:1991}. The
idea of building a laser upon this gain followed soon after its
first observation \cite{Hilico:1992}. Nevertheless, cold atoms in
optical cavity have been then mainly used for quantum optics
purposes \cite{Kimble:2005} and laser demonstrations with
different gain mechanisms are much more recent
\cite{Kruse:2003,Guerin:2008}.

If one could combine both gain and radiation trapping at the same
time in a cold atom cloud, this could give rise to a diffusive
random laser, as predicted by Letokhov \cite{Letokhov:1968}. Since
his original ``photonic bomb'' prediction, great efforts have been
made to experimentally demonstrate this effect in different kinds
of systems
\cite{Gouedard:1993,Lawandy:1994,Cao:1998,Cao:1999,Wiersma:2001,Strangi:2006,Gottardo:2008},
as well as to understand the basic properties of random lasing
\cite{Wiersma:1996,Burin:2001,Vanneste:2007,Tureci:2008}. The
broad interest of this topic is driven by potential applications
(see \cite{Wiersma:2008} and references therein) and by its
connection to the fascinating subject of Anderson localization
\cite{Anderson:1958,Conti:2008}. State-of-the-art random lasers
\cite{Wiersma:2008,Cao:2003,Cao:2005,Noginov} are usually based on
condensed matter systems, and feedback is provided by a disordered
scattering medium, while gain is provided by an active material
lying in the host medium or inside the scatterers. In general,
scattering and gain are related to different physical entities.

The peculiarity of a random laser based on cold atoms would be
that the same microscopic elements (the atoms) would provide both
ingredients (scattering and gain) of random lasing. On the one
hand, it leads to an easier characterization and modelling of the
microscopic properties of the system, which can be extremely
valuable for a better understanding of the physics of random
lasers. Moreover, cold atoms are ``clean'' and perfectly
characterized samples. In addition, relaxation of optical
coherences is limited by radiative processes, which makes the
transition to cooperative emission, such as superfluorescence
\cite{Bonifacio:1975,Paradis:2008}, more accessible than in
condensed matter systems. On the other hand, it is clear that
pumping atoms to induce gain reduces drastically their scattering
cross-section, due to the saturation effect
\cite{Cohen:PetitLivreRouge-EN}: atoms spend less time in their
ground state, in which they can scatter light. It is thus not
obvious at all that reasonable conditions for random lasing can be
obtained in cold atoms. Moreover, these conditions are expected to
be different for each gain mechanism.

The purpose of this article is to present the status of our
experimental and theoretical investigations on this issue. In the
next section we present our work on lasing with cold atoms, with a
standard cavity \cite{Guerin:2008}. Experimentally, this
demonstration of a cold-atom-based laser with different gain
mechanisms is a first important step towards the building of a
random laser. More fundamentally, this experiment can be compared
to theory to validate or improve the modelling of such gain media,
including saturation effect, laser dynamics or statistics. We
present a part of this modelling in section \ref{sect.model}.
Ultimately, when studying the random laser properties, it will be
of first importance to be able to discriminate which behaviour
originates from the particular used gain medium and mechanism and
which one is specific to the feedback mechanism (standard cavity
or scattering). The cavity-laser thus serves as a reference, to be
compared with theory and with the (forthcoming) random laser.

The second important step of this project, presented in section
\ref{sect.threshold}, is to quantify the constraint of combining
gain and scattering at the same time. This leads to the evaluation
of the random laser threshold \cite{Froufe:2009,Guerin:2009}. The
goal is first to establish the feasibility of a random laser with
cold atoms, which was not obvious, and second to compare the
different gain mechanisms in order to experimentally choose the
more appropriate one with the best pumping parameters. This is
necessary before conducting further experimental efforts to
achieve the threshold. As the results are promising, we draw some
prospects for our future work in the last section.

\section{Gain and lasing with cold atoms: experimental investigation}\label{sect.laser}

\begin{figure}[h]
\centering
\includegraphics{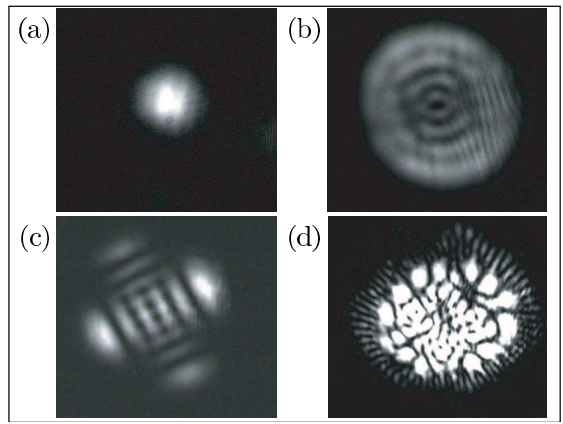}
\caption{Transverse modes of cold-atom lasers. (a) Gaussian
TEM$_{00}$ mode, obtained by inserting a small diaphragm in the
cavity. Typical modes of: (b) the Mollow laser, (c) the Raman
laser, and (d) the four-wave mixing laser respectively.}
\label{fig.lasers}
\end{figure}

In this section, we briefly describe our experimental setup and
then the different gain mechanisms we have studied. Mollow gain is
the simplest as it involves only one pumping field and two-level
atoms \cite{Mollow:1972,Wu:1977}. By using the more complex atomic
structure of rubidium atoms, we can create two-photon transitions
between two non-degenerate ground states. This can produce Raman
gain \cite{Tabosa:1991,Grison:1991,Hilico:1992}. Finally, the
atomic non-linearity can give rise to parametric gain, for example
with four-wave mixing. We have demonstrated laser action with each
of these mechanisms (figure \ref{fig.lasers}) \cite{Guerin:2008}.
We mention also briefly some other gain mechanisms that could be
used.

\subsection{Experimental setup}\label{sect.setup}

Our experiment uses a cloud of cold $^{85}$Rb atoms confined in a
vapour-loaded Magneto-Optical Trap (MOT) \cite{Metcalf} produced
by six large independent trapping beams, allowing the trapping of
up to $10^{10}$ atoms at a density of $10^{10}$~atoms/cm$^3$,
corresponding to an on-resonance optical thickness $b_0 \sim 10$.
A linear cavity, formed by two mirrors separated by a distance
$L=0.8$~m, is placed outside the vacuum chamber. Reflections on
the vacuum cell yields a low finesse $\mathcal{F}=16$. To add gain
to our system, we use either one or two counter-propagating pump
beams, denoted F (forward) and B (backward), produced from the
same laser with a waist $w_\mathrm{pump}=2.6$~mm, with linear
parallel polarizations and a total available power $P=80$~mW,
corresponding to a maximum pump intensity $I=2P/(\pi
w_\mathrm{pump}^2)\approx\,750$~mW/cm$^2$. The pump is tuned near
the $F=3\rightarrow F'=4$ cycling transition of the $D2$ line of
$^{85}$Rb (frequency $\omega_0$, wavelength $\lambda = 780$~nm,
natural linewidth $\Gamma/2\pi = 6.1$~MHz), with an adjustable
detuning $\Delta=\omega_\mathrm{F,B}-\omega_0$ and has an incident
angle of about $20^\circ$ with the cavity axis. An additional beam
P is used as a local oscillator to monitor the laser spectrum or
as a weak probe to measure transmission or reflection spectra
(figures \ref{fig.Raman}(b) and \ref{fig.FWM}(b)) with a
propagation axis making an angle with the cavity axis smaller than
$10^\circ$. Its frequency $\omega_\mathrm{P}$ can be swept around
the pump frequency with a detuning
$\delta=\omega_\mathrm{P}-\omega_\mathrm{F,B}$. Both lasers, pump
and probe, are obtained by injection-locking of a common master
laser, which allows to resolve narrow spectral features. In our
experiments, we load a MOT for 29~ms, and then switch off the
trapping beams and magnetic field gradient during 1~ms, when
lasing or pump-probe spectroscopy are performed. In order to avoid
optical pumping into the dark hyperfine $F=2$ ground state, a
repumping laser is kept on all time.

\subsection{Gain mechanisms}

Contrary to most laser gain media, cold atoms do not present
non-radiative, fast-decaying transitions, preventing a standard 4-level scheme to produce a population inversion, as e.g. in many gas lasers.
Nevertheless there are many different mechanisms which allow an inversion between two different atomic states.
These can either be different states in external degrees of freedom (momentum or vibrational levels in an external potential) or internal degrees of freedom (dressed states \cite{Cohen:PetitLivreRouge-EN} or different ground states).
We have used two mechanisms, which we call respectively Mollow and
Raman gain (described in the two next subsections), to induce
lasing with a cavity and to study the threshold of a random laser
(section \ref{sect.threshold_evaluation}). The atomic nonlinearity
can also be used to obtain parametric gain and lasing (subsection
\ref{sect.FWM}), as well as other, more complicated, schemes using
quantum interferences or the external degrees of freedom, as
briefly mentioned in subsection \ref{sect.othergain}.

\subsubsection{Mollow gain}\label{sect.mollowgain}

\begin{figure}[t]
\centering
\includegraphics{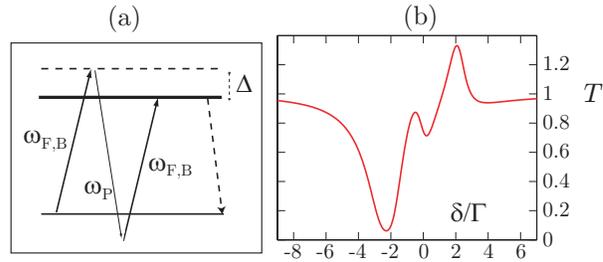}
\caption{(a) Principle of the Mollow gain depicted as a
three-photon transition from the ground state to the excited
state. It can also be viewed as a population inversion in the
dressed-state basis (figure \ref{dynamics}(a)). (b) Transmission
spectrum, computed for typical experimental parameters $b_0 = 10$,
$\Omega = 2 \Gamma$ and $\Delta = \Gamma$.}\label{fig.Mollow}
\end{figure}

The most simple gain mechanism we can imagine in cold atoms was
described by Mollow \cite{Mollow:1972} and observed soon
afterwards \cite{Wu:1977} with atomic beams. It involves a
two-level atom driven by one strong pumping field. The driving
field induces a population inversion in the dressed-state basis
\cite{Cohen:PetitLivreRouge-EN} (see figure \ref{dynamics}(a)) and
therefore a weak probe beam can then by amplified. The whole
process can also be described in the bare-states basis by a
three-photon transition from the ground state to the excited state
via two absorptions of pump photons  as sketched in figure
\ref{fig.Mollow}(a).

The main amplification feature appears for a pump-probe detuning
$\delta = \mathrm{sign}(\Delta) \sqrt{\Delta^2 + \Omega^2}$, where
$\Omega$ is the Rabi frequency of the pump-atom coupling, related
to the pump intensity $I$ by $\Omega^2 = \mathcal{C}^2 \Gamma^2
I/(2I_\mathrm{sat})$ ($I_\mathrm{sat}=1.6$~mW/cm$^2$ is the
saturation intensity and $\mathcal{C}$ is the averaged squared
Clebsch-Gordan coefficient of the $F=3\rightarrow F'=4$
transition) and has a typical width on the order of $\Gamma$. Note that
another, dispersion-like feature appears around $\delta =0$, which
is associated with two-photon spontaneous emission processes
\cite{Grynberg:1993}. This contribution also induces gain but with
a much smaller amplitude. Note that this can generate lasing
without population inversion
\cite{Grandclement:1987,Zakrzewski:1992_2,Mompart:2000}.

\begin{figure}[b]
\centering
\includegraphics{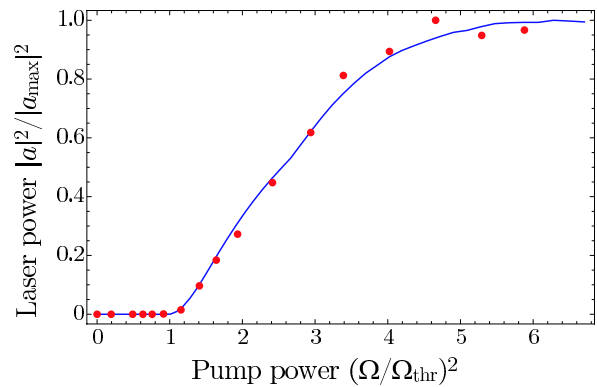}
\caption{Laser power versus pump power. Points are experimental
data obtained after an average of 1000 experiments, the continuous
line is the theoretical fit (see section \ref{sect.model}) with a
single fit parameter $\eta \simeq 8$ (corresponding to $N \sim
10^8$ atoms in the cavity mode) and all other parameters fixed at
their experimental values, $b_0 = 11$, $\Delta = \Gamma$,
$\mathcal{F} = 16$, $L = 0.8$~m, and cavity waist at the MOT
location $w = 500~\mu$m. The horizontal axis is normalized by the
threshold value of the squared Rabi frequency
$\Omega_{\mathrm{thr}}^2$; the vertical axis is normalized by the
maximum value of the electromagnetic field intensity reached at
about $(\Omega/\Omega_{\mathrm{thr}})^2 \sim 5$.}
\label{fig.puissMollow}
\end{figure}

In our experiment, we have measured single-pass gain as high as
50~\%, which is more than enough to induce lasing even with a
low-finesse cavity \cite{Guerin:2008}. This Mollow laser has an
output intensity reaching 35~$\mu$W achieved for $|\Delta|\sim
2\Gamma$. Its threshold in pump intensity is in
agreement with the corresponding measured single-pass gain and the losses
of the cavity. The laser polarization is linear, parallel to the
pump polarization, because this is the configuration for which
gain is maximum, as the driven atomic dipole is then parallel to
the probe field. We have also measured the output power as a
function of the pump intensity, as reported in figure
\ref{fig.puissMollow}. We observe a threshold and the maximum
intensity is reached for a pump power about 5 times larger than
at the threshold. This behaviour is well described by the
theoretical model presented in section \ref{sect.model}.

The Mollow laser works for pump detuning $|\Delta|<4\Gamma$. When
the pump frequency in detuned farther away from the atomic
resonance, Raman gain becomes dominant, and the system switches to
another regime of laser, based on Raman gain.

\subsubsection{Raman gain}\label{sect.ramangain}

\begin{figure}[b]
\centering
\includegraphics{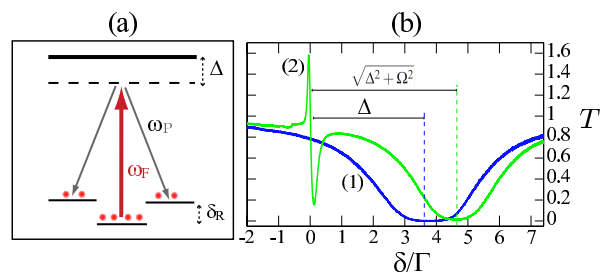}
\caption{(a) Principle of the Raman mechanism, depicted here for a
$F=1 \rightarrow F'=2$ transition. (b) Experimental transmission
spectra, plotted as a function of the pump-probe detuning
$\delta$. Without pumping, spectrum (1) shows only the atomic
absorption. A pump beam of detuning $\Delta=-3.8~\Gamma$ and
intensity $13$~mW/cm$^2$, corresponding to a Rabi frequency
$\Omega = 2.5~\Gamma$, is added to obtain spectrum (2), which then
exhibits a Raman resonance in the vicinity of $\delta=0$. Moreover, the
atomic absorption is shifted due to the pump-induced light shift
and the absorption is reduced due to saturation.}\label{fig.Raman}
\end{figure}

Raman transitions refer in general to two-photons transitions
between two non-degenerate ground states, the intermediate energy
level being in the vicinity of the atomic excited states. To
obtain gain, a pumping field can induce the first upward
transition and a probe beam can then be amplified at the frequency
of the downward transition. In our case, Raman gain relies on the
pump-induced population inversion among the different
light-shifted $m_F$ Zeeman sublevels of the $F=3$ hyperfine level
\cite{Tabosa:1991,Grison:1991}, as depicted in figure
\ref{fig.Raman}(a). The optical pumping induced by the
$\pi$-polarized pump laser leads to a symmetric distribution of
population with respect to the $m_F=0$ sublevel of the ground
state, with this sublevel being the most populated and also the
most shifted, due to a larger Clebsch-Gordan coefficient
\cite{Brzozowski:2005}. To record a transmission spectrum, atoms
are probed with a $\pi$-polarized (with perpendicular direction)
probe beam, thus inducing $\Delta m_F = \pm 1$ Raman transitions.
Depending on the sign of the pump-probe detuning $\delta$, the
population imbalance induces gain or absorption. Each pair of
neighboring sublevels contributes with a relative weight depending
on the population inversion. In practice however, the
contributions of different pairs are not resolved and only two
structures (with opposite signs) are visible, one corresponding to
amplification for $\delta=-\delta_\mathrm{R}$ and one to
absorption for $\delta=\delta_\mathrm{R}$. Note that this
situation corresponds to a red detuning for the pump ($\Delta <
0$) and that the signs are inverted for blue-detuning ($\Delta >
0$). As $\delta_\mathrm{R}$ comes from a differential light-shift
(because of different Clebsch-Gordan coefficients), it is usually
on the order of $\Gamma/10$, whereas $\Delta$ is a few $\Gamma$.
The width $\gamma$ of the resonances is related to the elastic
scattering rate, also much smaller than $\Gamma$
\cite{Grison:1991}. Far from the main atomic absorption resonance,
the Raman resonance is thus a narrow spectral feature, as in
figure \ref{fig.Raman}(b)

The laser obtained with Raman gain has an output polarization
orthogonal to the pump one (contrary to the Mollow laser) and less
power (2~$\mu$W). Moreover, the sharpness of the gain curve makes
the Raman gain very sensible to any Doppler shift. The radiation
pressure from the pump beam makes thus the laser emission to stop
after only $\sim 20~\mu$s. On the other hand, the narrow spectrum
of the laser can be easily characterized by a beat-note experiment
\cite{Guerin:2008}.

\subsubsection{Parametric gain}\label{sect.FWM}

The Mollow and Raman lasers only require one pump beam. By using a
second pump beam, we can induce four wave mixing (FWM): the two
pumps of frequencies $\omega_\mathrm{F}$ and $\omega_\mathrm{B}$
and one probe -- or initial fluctuation -- of frequency
$\omega_\mathrm{P}$ generate a fourth field at frequency
$\omega_\mathrm{C}$, called the conjugate field
\cite{Yariv:1977,Abrams:1978,Boyd:1981}. The frequencies and
wave-vectors of all the fields are related by energy and momentum
conservations. If we want to obtain \emph{gain} for the probe, we
have to choose a configuration where the conjugate frequency
equals the probe one: $\omega_\mathrm{C}=\omega_\mathrm{P}$. Then,
the pump frequencies have to fulfill the condition
$\omega_\mathrm{F} + \omega_\mathrm{B} = 2 \omega_\mathrm{P}$.
From an experimental point of view, the most simple configuration
consists of all frequencies to be the same (``degenerate FWM'').
This is the experimental situation that we have studied so far,
and we did obtain lasing in cold atoms with this mechanism
\cite{Guerin:2008}. Note that this mechanism has been observed a
long time ago with hot atoms
\cite{Kleinmann:1985,Pinard:1986,Leite:1986}.

\begin{figure}[b]
\centering
\includegraphics{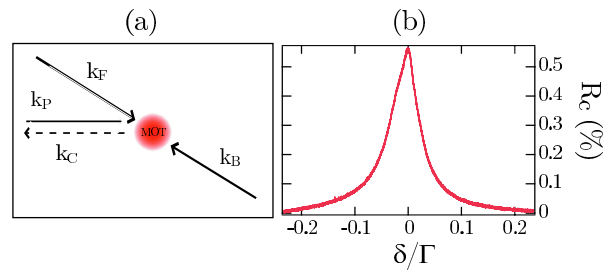}
\caption{(a) Principle of four-wave mixing. (b) Typical
experimental reflection spectrum.}\label{fig.FWM}
\end{figure}

Due to the phase matching condition, however, the gain is not in
the forward transmission of the probe beam, but in backward
reflection, provided that the two pumps are counterpropagating.
The conjugate beam is actually the phase-conjugate of the probe
beam. This property has a number of consequences for the laser
\cite{Guerin:2008}. First, it leads to a different threshold
condition \cite{Pinard:1986}: a reflectivity of only 1~\% is
enough to generate lasing, despite the much larger losses of the
cavity (32~\% for a round trip). This is due to constructive
interferences between transmitted and reflected waves, as observed
in double-pass experiments \cite{Michaud:2007}. Second, it leads
to more complex transverse modes (figure \ref{fig.lasers}(d)),
because the phase conjugation mechanism allows any transverse
pattern to be stable through the resonator \cite{Lind:1981}.

Finally, these properties lead to a much larger power than with
Mollow and Raman gain, as up to $300~\mu$W have been obtained.
Moreover, this laser might find application for other topics like
pattern formation \cite{Grynberg:1994,Ackemann:2001} or the
production of twin beams for quantum optics
\cite{Vallet:1990,McCormick:2007} and even quantum imaging
\cite{Boyer:2008}. Note also that the use of two pump beams allows
a longer laser emission (figure \ref{fig.dynamicsExp}(b)) because
the effect of radiation pressure on the cloud is suppressed (other
effects can arise due to the dipole force, see
\cite{Gattobigio:2006}).

\subsubsection{Other gain mechanisms}\label{sect.othergain}

Our previous study on the gain mechanisms that can be used in cold
atoms is not exhaustive. For example, the two hyperfine ground
states of rubidium atoms can also be used to produce Raman gain
\cite{Kumar:1985}.

Other, more complicated, schemes, involve quantum interference to
induce gain without population inversion (whatever the basis)
\cite{Mompart:2000}. This can be realized with a $\Lambda$ scheme
\cite{Padmabandu:1996} or a V scheme
\cite{Zibrov:1995,Kitching:1999}. In this last configuration, a
large detuning between the pump and the gain frequency can be
reached by using the two D lines of rubidium, whose separation is
15~nm. If used to produce a random laser, this can be highly
valuable to make the detection easier (see the discussion of
section \ref{sect.conclusion}).

Another possibility is to use the atomic external degrees of
freedom, i.e., their kinetic energy. Transitions between different
velocity classes produce recoil-induced resonances
\cite{Courtois:1994}, and high gain can be achieved
\cite{Vengalattore:2005}. These resonances can ultimately lead to
a ``Collective atomic recoil laser''
\cite{Bonifacio:1994,Berman:1999}, which has been demonstrated
with cold atoms \cite{Kruse:2003}.

Finally, one could also consider higher-order photonic processes,
such as two-photon dressed-state lasers \cite{Gauthier:1992}.

\subsection{Laser dynamics}\label{sect.dynamics}

We have been interested so far only in steady-state properties.
Our cold-atom-based laser could also reveal
interesting dynamical properties.

\begin{figure}[b] 
\centering
\includegraphics{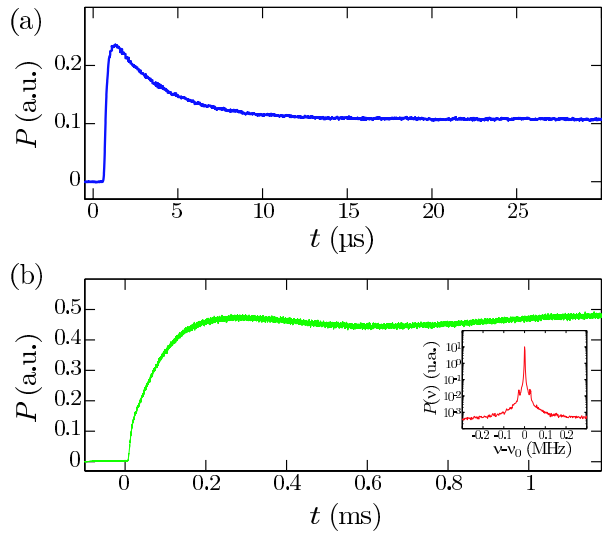}
\caption{Temporal evolution of the Mollow (a) and FWM lasers (b).
The pump beams are switched on at $t=0$. Note the change in the
time scales. The Mollow laser switches on very fast (the initial
decrease is due to optical pumping to the dark hyperfine state),
whereas the FWM laser needs more time. Inset of (b): Power
spectrum, with a logarithmic scale, of the FWM laser measured by a
beating with a local oscillator of frequency $\nu_0$. All data are
the result of an average of 1000 cycles.} \label{fig.dynamicsExp}
\end{figure}

For example, we have observed that the Mollow laser switches on
very quickly, whereas the FWM laser switches on very slowly
(figure \ref{fig.dynamicsExp} -- notice the change in the time
scales). The laser rise-times are related to the spectral
properties of the gain mechanism. We are indeed in the bad cavity
limit, where the decay rate of the cavity is not smaller than the
spectral width of the amplification. The measured orders of
magnitude ($\sim 0.2~\mu$s for the Mollow laser, more than
$100~\mu$s for the FWM laser) are consistant with the inverse of
the laser spectral width (a few MHz for the Mollow laser, a few
kHz for the FWM laser). Note that the spectral width of the FWM
laser, measured by a beating with a local oscillator (inset of
figure \ref{fig.dynamicsExp}(b) -- note the logarithmic scale), is
substantially smaller than the width of the reflectivity spectrum
(figure \ref{fig.FWM}(b)).

Clearly, further experimental and theoretical studies would be
needed to fully understand the dynamical behaviour of these
different lasers. However, a first insight is given in the next
section.


\section{Model of the Mollow laser}\label{sect.model}

We present in this section a theoretical analysis of our
cold-atom laser based on Mollow gain.

\subsection{Formalism}

We consider an ensemble of $N$ identical two-level atoms pumped by
a strong field of frequency $\omega_L$, detuning $\Delta$ and
intensity characterized by the Rabi frequency $\Omega$. They are
placed inside a cavity of eigenfrequencies $\omega_m$, and $a_m$,
$a_m^{\dagger}$ are the annihilation and creation operators,
corresponding to the mode $m$. The cavity is partly open and the
energy of the electromagnetic field can escape at a rate $2
\kappa$. We recall that we are in in the bad cavity limit
\cite{Zakrzewski:1992_1}.

To describe the dynamics of the coupled system of atoms and modes
of the electromagnetic field, we use the master equation approach
\cite{Cohen:PetitLivreRouge-EN,Zakrzewski:1992_1,Zakrzewski:1991}.
The density matrix operator obeys
\begin{equation}
\overset{.}{\rho}=-i\left[H,\rho\right] + L_{A}\rho + L_{F}\rho \;
, \label{master}
\end{equation}
where we set $\hbar = 1$ and the Hamiltonian is
\begin{equation}
H=\frac{1}{2}\sum_{j=1}^{N}\left[ -\Delta \cdot
\sigma_{3j}+\Omega\left(\sigma_{j}+\sigma_{j}^{{
\dagger}}\right)+\sum_{m}^{M}\left(g_{jm}\sigma_{j}^{{
\dagger}}a_{m}+h.c.\right)\right]
+\sum_{m=1}^{M}\delta_{m}a_{m}^{\dagger}a_{m}\; . \label{ham}
\end{equation}
Here the rotating-wave approximation was used and the operators $\sigma_{3j}$, $\sigma_{j}^{\dagger}$ and
$\sigma_{j}$ are standard Pauli matrices, describing the populations ($\sigma_{3j}$) and the coherences ($\sigma_{j}$) of the
$j$-th atom. The detuning of the mode frequency from the pump frequency is $\delta_m$ and $g_{jm}$ are the coupling constants between atoms and cavity
modes, which we assume to be independent of the mode index $m$ and to have the same
absolute value $g$ for all atoms and random phases $\phi_j$:
$g_{jm} = g\exp(i\phi_{j})$. In the Hamiltonian of equation
(\ref{ham}), only a part of modes of the electromagnetic field in
the cavity ($M$ modes) are treated explicitly, whereas spontaneous
emission into other modes is taken into account through the first
non-Hamiltonian term of the master equation (\ref{master}):
\begin{eqnarray}
L_{A}\rho &=& \frac{\Gamma}{2}\sum_{j=1}^{N}
\left(2\sigma_{j}\rho\sigma_{j}^{{ \dagger}}-\sigma_{j}^{{
\dagger}}\sigma_{j}\rho-\rho\sigma_{j}^{{
\dagger}}\sigma_{j}\right) \; .\label{eq.La}
\end{eqnarray}
The second non-Hamiltonian term in equation (\ref{master})
describes cavity damping of the electromagnetic modes:
\begin{eqnarray}
L_{F}\rho &=& \kappa\sum_{m=1}^{M} \left(2a_{m}\rho a_{m}^{{
\dagger}}-a_{m}^{{ \dagger}}a_{m}\rho-\rho a_{m}^{{
\dagger}}a_{m}\right) \; .\label{eq.Lf}
\end{eqnarray}

\begin{figure}[t]
\centering
\includegraphics{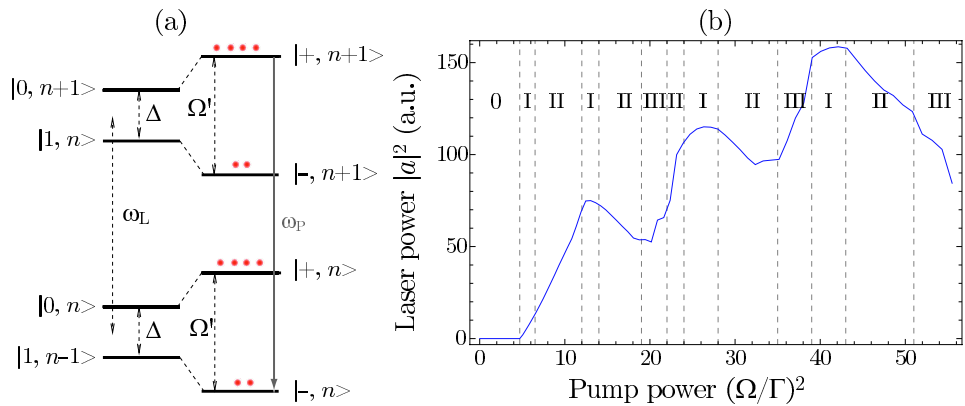}
\caption{(a) Coupling of the bare atomic states $|0,n\rangle$,
$|1, n\rangle$ (left) to the dressed states $|-, n\rangle$, $|+,
n\rangle$ (right), where $n$ stands for the number of pump
photons. The population inversion and the subsequent gain is
sketched on the dressed-state representation. (b) Laser power
versus pump power for a laser with a single cavity mode
interacting with atoms ($M = 1$). Values of parameters are chosen
as in the experiment, $b_0 = 11$, $\Delta = \Gamma$, $\mathcal{F}
= 16$, $L = 0.8$~m, $w = 500~\mu$m, and we have taken $\delta_m =
1.8 \Gamma$ and $\eta \simeq 15$, corresponding to $N = 2.3 \times
10^8$. Regions I, II and III correspond to three different types
of behaviour illustrated in figure \ref{graph6}. Regions III
(chaotic behaviour) appear only when $\eta$ is large enough.
(e.g., no chaotic behaviour is observed for $\eta = 8$ as in
figure \ref{fig.puissMollow}).} \label{dynamics}
\end{figure}

The physical processes described by the set of equations
(\ref{master}--\ref{eq.Lf}) become especially clear in the
dressed-state basis \cite{Cohen:PetitLivreRouge-EN}. The dressed
states are eigenstates of the system ``atoms + pump field''. In
the dressed-state basis, the states $\left|1\right\rangle_{j}$ and
$\left|0\right\rangle_{j}$ of the two-level atom $j$ become
$\left|+\right\rangle _{j}=(\cos\theta)\left|1\right\rangle
_{j}+(\sin\theta)\left|0\right\rangle _{j}$ and
$\left|-\right\rangle _{j}=-(\sin\theta)\left|1\right\rangle
_{j}+(\cos\theta)\left|0\right\rangle _{j}$, respectively, where
 $\Omega=\Omega'\sin2\theta$ and $\Delta = -\Omega'\cos2\theta$
with $\Omega'=\sqrt{\Omega^{2}+\Delta^{2}}$ being the generalized
Rabi frequency (figure \ref{dynamics}(a)). Equations (\ref{ham})
and (\ref{eq.La}) can then be transformed (equation (\ref{eq.Lf})
remains unchanged) and equation (\ref{master}) leads to the
following semi-classical equations for quantum-mechanical
expectation values of operators $\sigma_{j}$,
$\sigma_{3j}$ and $a_{m}$
\cite{Zakrzewski:1992_1,Zakrzewski:1991}:
\begin{eqnarray}
\overset{.}{\sigma}_{3j} & = &
-\gamma_{1}\left(\sigma_{3j}-\bar{\sigma}_{3j}\right)+ \Phi(\sigma_j,a_m) \; ,
\label{eq.pop}
\\
\overset{.}{\sigma}_{j} & = & -\left(\gamma_{2}+i\Omega'
\right)\sigma_{j} + \Psi(\sigma_j, \sigma_{3j}, a_m) \; ,
\label{eq.coh}
\\
\overset{.}{a}_{m} & = &
-\left(\kappa+i\delta_{m}\right)a_{m} + \Theta(\sigma_{3j},\sigma_j) \; ,
\label{eq.champ}
\end{eqnarray}
where $\bar{\sigma}_{3j} = -2\cos2\theta/(1+\cos^{2}2\theta)$,
$\gamma_{1}=(\Gamma/2)\left(1+\cos^{2}2\theta\right)$,
$\gamma_{2}=(\Gamma/4)\left(2+\sin^{2}2\theta\right)$ and the
expressions of functions $\Phi$, $\Psi$, $\Theta$ are given in
\ref{appendix}. The equations for $\sigma_j^*$ and $a_m^*$ are
complex conjugates of equations (\ref{eq.coh}) and
(\ref{eq.champ}), respectively. These equations describe
respectively the evolution of the populations, the coherences and
the cavity field. The first term of each right-hand-side contains
the natural evolution associated with rates $\gamma_1$, $\gamma_2$
and $\kappa$ as well as the driving by the pump field (Rabi
frequency $\Omega$). The equations are coupled by their second
r.h.s. terms (see equations \ref{eq.Phi}--\ref{eq.Theta} for the
complete expressions), which contain the atom-field coupling $g$
and the number $N$ of atoms coupled to the cavity mode. For a
Gaussian cavity mode with waist $w$ one obtains \cite{mu:1992} $g
\simeq \sqrt{3c\Gamma/L}/k_0 w$, $N \simeq (k_0 w)^2 b_0/12$ where
$k_0 = \omega_0/c = 2 \pi/\lambda$, $b_0$ is the optical thickness
of the cloud at resonance without pump and $L$ the length of the
cavity. To take into account the fact that volumes and shapes of
the atomic cloud and cavity modes are different and not Gaussian,
we will multiply $N$ by a free parameter $\eta$ that will be
adjusted to fit experimental results.

Equations (\ref{eq.pop}--\ref{eq.champ}) can be solved numerically
following the procedure described in \ref{appendix}.

\subsection{Dynamics of the laser}

\begin{figure}[t]
\centering
\includegraphics[width=16cm]{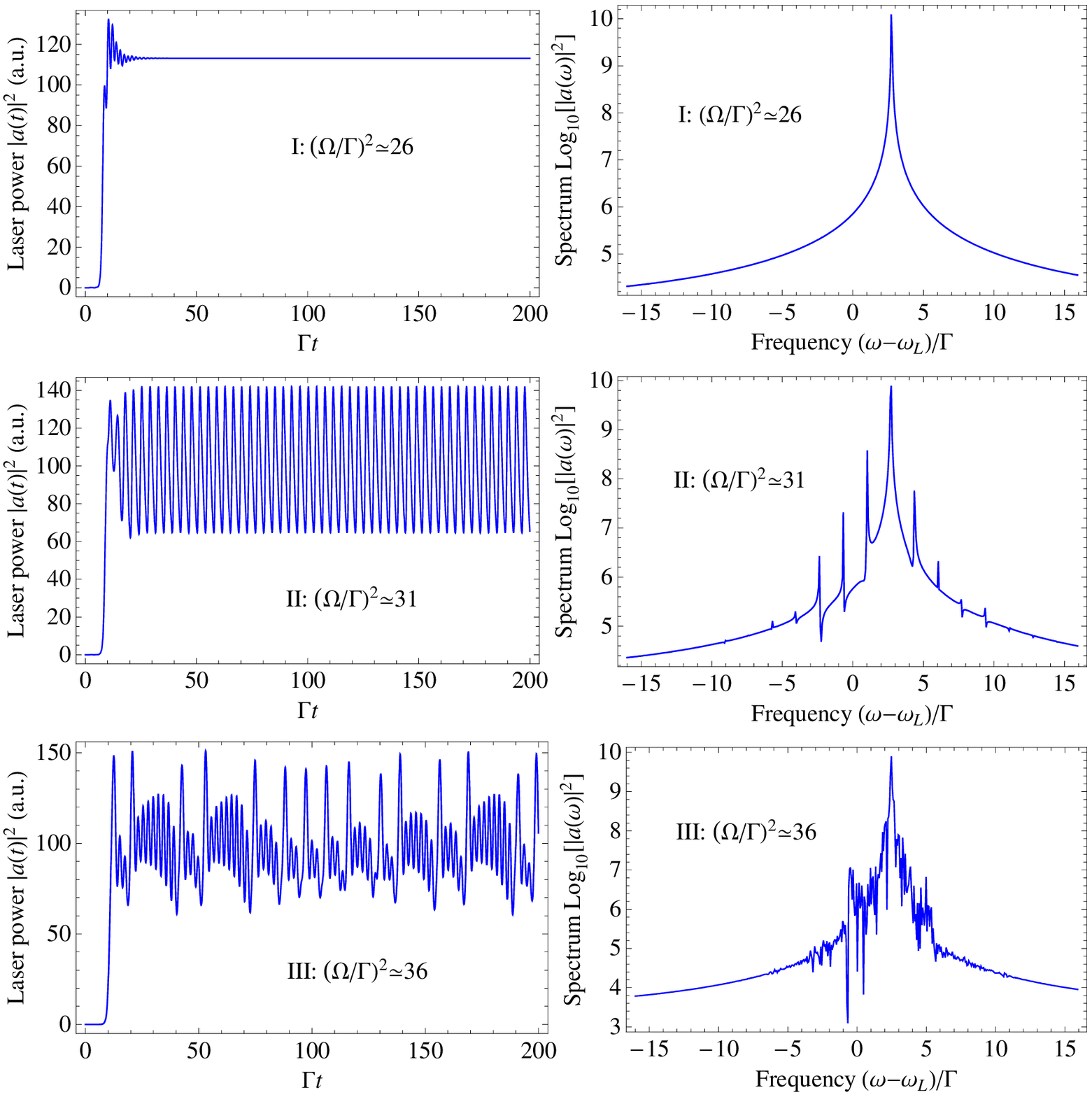}
\caption{Laser power versus time (left) and corresponding spectra (right) for the three different regimes of lasing: I --
single-mode lasing, II -- multi-mode lasing, and III -- chaotic
behaviour. The spectra are calculated in the stationary regime
$\Gamma t > 100$. The parameters used in the calculation are the
same as in figure \ref{dynamics}.} \label{graph6}
\end{figure}

The dynamics of the laser depends strongly on how the different
time scales associated with the evolutions of populations,
coherences and field can be compared. Following the classification
introduced in \cite{Arecchi:1984,Tredicce:1984}, class A lasers
are those in which population and polarization decay much faster
than the field so that the dynamics is governed by the single
field equation (for example He-Ne, Ar+, Kr+, dye lasers), class B
lasers are those for which the population decays slowly and two
equations are necessary (e.g. ruby, Nd, CO$_2$ lasers) and
finally, class C lasers are those for which the three time scales
are of the same order of magnitude. The last ones are known to
exhibit chaotic behaviour \cite{Arecchi:1984,Tredicce:1984}.

For our cold atom sample, the relaxation rate of atomic coherences
is only given by optical processes (no collision, e.g.). As a
consequence, and with Mollow gain, the time scales for the
evolution of the populations and coherences are both on the order
of $\Gamma^{-1}$ (note that this is not the case with Raman gain
or FWM). The cavity damping rate $\kappa$ is related to the
finesse $\mathcal{F}$ of the cavity by $\kappa = \pi c/2 L
\mathcal{F}$. With our experimental parameters, it gives
$\kappa/2\pi \simeq 5.9$~MHz, i.e., $\kappa \sim \Gamma$ and we
are thus in a regime corresponding to class C lasers. This is
consistent with the behaviour observed in our numerical
simulation, as described in figures \ref{dynamics}(b) and
\ref{graph6}. We show the dependence of the laser power on the
pump power in figure \ref{dynamics}(b) for the case of when only a
single mode ($M = 1$) with a well-defined frequency (detuning
$\delta  = 1.8\Gamma$) is present. The non-monotonic character of
this dependence is accompanied by a complex dynamic behaviour that
we illustrate in figure \ref{graph6}. Three distinct dynamic
regimes are possible, depending on the pump strength: the laser
emits at a single frequency (I) or at several frequencies at a
time (II). In addition, chaotic behaviour appears for certain
$\Omega$ (III). These results show that Mollow gain, associated to
a low-finesse cavity, is perfectly suited for the study of chaotic
regimes in lasers, as already predicted in
\cite{Zakrzewski:1992_1}.

We did not observe this kind of dependence of laser power as a
function of the pump intensity in our experiment. This behaviour
appears indeed only for some values of atom number or detuning
$\delta_m$ and a specific experimental implementation would be
needed. To model the experimental results, we solved the coupled
laser equations with a well-defined single mode of the field ($M =
1$, detuning $\delta_1 = \delta$) and average the resulting laser
intensity $|a|^2$ over 200 equidistant $\delta$ between $-2
\Gamma$ and $5 \Gamma$. This is supposed to account for the fact
that the exact frequency of the cavity mode is not precisely known
in the experiment and fluctuates shot to shot, as well as for
averaging over many consecutive measurements (1000 in the
experiment). Using $\eta$ as a free parameter, we are then able to
obtain a good fit to experimental data (solid line in figure
\ref{fig.puissMollow}). The obtained value of $\eta = 8$ is in
qualitative agreement with the observation of high-order
transverse laser modes: the corresponding cavity modes have a
larger overlap with the atom cloud compared to a gaussian
TEM$_{00}$ mode.

\section{Threshold of a random laser with cold atoms}\label{sect.threshold}

We now turn to the more specific question of realizing a
\emph{random} laser based upon one of the gain mechanisms studied
in the previous section. As explained in the introduction, gain is
not the only necessary ingredient for building a random laser, as
scattering should also be present to provide feedback. In the
previous laser experiment, scattering was not considered, and the
optimum conditions for the laser operation are those which
optimize the single-pass transmission though the cloud, which are
very probably not the one that optimize scattering. To establish
the best conditions for random lasing, we thus have to quantify
more precisely the constraint of combining gain and scattering.
This condition is already contained in the seminal Letokhov's
paper \cite{Letokhov:1968}, and leads indeed to the evaluation of
the random laser threshold.

\subsection{Random laser threshold}

Although different kinds of random lasers have been achieved
experimentally, several points are still under debate. Random
lasers may be divided into two categories, depending on the
feedback mechanism~\cite{Cao:2005}. In the regime of incoherent
(or intensity) transport,
 feedback is provided by an increase of the photon path lengths (or lifetime) in the system.
 This regime has been observed experimentally in many systems, such as
 powdered laser crystals~\cite{Gouedard:1993}, dye solutions containing
 microparticles~\cite{Lawandy:1994}, solid-states materials~\cite{Noginov},
 or porous glass infiltrated by a dye solution in liquid crystal~\cite{Wiersma:2001}.
 In a regime in which interference effects survive the random scattering process,
 a coherent (or field) feedback is expected, leading to a behaviour closer
 to that taking place in conventional lasers. This behaviour is expected to become
 substantial close to the Anderson localization regime \cite{Anderson:1958}. Experiments in this direction
 have been carried out, e.g., on zinc oxide powders~\cite{Cao:1999,Cao:2005}, and
 signatures of coherent feedback on the random lasing mechanism have been
discussed recently~\cite{Vanneste:2007,Tureci:2008,Conti:2008}.
For a given practical realization of a random laser,
discriminating between a coherent or incoherent feedback mechanism
is a complicated task. For example, narrow emission peaks above
threshold can be observed, even far from the localization
threshold~\cite{Cao:2005,Wiersma:2004}. The existence of such
peaks in experiments can be attributed to exponential gain along
very long diffusion paths~\cite{Wiersma:2004}, or to interferences
and coherent feedback in the weak scattering
limit~\cite{Vanneste:2007}. On the route towards a random laser
with cold atoms, we first consider a diffusive random laser (with
intensity feedback). This approach is motivated by the fact that
experiments that should be carried out in the near future are
expected to work in this regime. Another motivation is the
identification of specific signatures of the incoherent feedback
regime that could guide the experimental observations. A deviation
from these signatures could be a measure of the onset of a
coherent feedback mechanism.

\subsection{Letokhov's threshold for cold
atoms}\label{sect.letokhov}

From Letokhov's diffusive description of light transport in a
homogeneous, disordered and active medium of size $L$, we know
that the random laser threshold is governed by two characteristic
lengths: the elastic scattering mean free path\footnote[7]{We
consider only isotropic scattering so that the transport length
equals the scattering mean free path.} $\ell_\mathrm{sc}$
\cite{Rossum:1999} and the linear gain length $\ell_\mathrm{g}$
($\ell_\mathrm{g}<0$ corresponds to absorption or inelastic
scattering). In the diffusive regime, defined as $L \gg
\ell_\mathrm{sc}$, the lasing threshold is reached when the
unfolded path length, on the order of $L^2/\ell_\mathrm{sc}$,
becomes larger than the gain length. More precisely, the threshold
is given by \cite{Letokhov:1968,Cao:2003}
\begin{equation}\label{eq.letokhov}
L_\mathrm{eff} > \beta \pi \sqrt{\ell_\mathrm{sc}\,
\ell_\mathrm{g} /3} \; ,
\end{equation}
where $\beta$ is a numerical factor that depends on the geometry
of the sample ($\beta=1$ for a slab, $\beta=2$ for a sphere, which
is the case we consider in the following), and $L_\mathrm{eff} =
\eta L$ is the effective length of the sample, taking into account
the extrapolation length \cite{Rossum:1999}. For $L >
\ell_\mathrm{sc}$ and a sphere geometry, $\eta = 1 + 2 \xi /
\left[L/\ell_\mathrm{sc} + 2\xi \right]$ with $\xi \simeq 0.71$
\cite{Zweifel,Drozdowicz:2003}. Note that deeply in the diffusive
regime ($L \gg \ell_\mathrm{sc}$), $\eta \sim 1$. Another
important length scale is the extinction length
$\ell_\mathrm{ex}$, as measured by the forward transmission of a
beam through the sample, $T = e^{-L/\ell_\mathrm{ex}}$. The
extinction length is related to the other lengths by
$\ell_\mathrm{ex}^{-1} = \ell_ \mathrm{sc}^{-1}
-\ell_\mathrm{g}^{-1}$. Note that this reasoning may not be
appropriate for backward gain such as produced by FWM.

For an atomic vapour, these characteristic lengths can both be
computed as a function of the atomic polarizability
$\alpha(\omega)$ at frequency $\omega$. The extinction
cross-section is indeed given by $\sigma_\mathrm{ex}(\omega) = k
\times \mathrm{Im}[\alpha(\omega)]$ and the elastic scattering
cross-section by $\sigma_\mathrm{sc}(\omega) = k^4/6\pi \times
|\alpha(\omega)|^2$ \cite{Lagendijk:1996} ($k=\omega/c$ is the
wave vector). Note that the first relation is general to any
dielectric medium whereas the second one is specific to dipole
scatterers. The characteristic lengths are then
$\ell_\mathrm{ex,sc}^{-1} = \rho\, \sigma_\mathrm{ex,sc}$, where
$\rho$ is the atomic density. The gain cross-section can be
defined the same way by $\ell_\mathrm{g}^{-1} =
\rho\,\sigma_\mathrm{g}$. The vapour is supposed at constant
density and homogenously pumped, so that both $\rho$ and $\alpha$
are position-independent. Even though this is not the precise
geometry of a cold-atom experiment, it allows us to perform
analytical estimations. As we consider resonant scatterers, we
deal only with quasi-resonant light and we shall use
$k=k_0=\omega_0/c$ with $\omega_0$ the atomic eigenfrequency. In
the following, we shall also use a dimensionless atomic
polarizability $\tilde{\alpha}$, defined as $\alpha =
\tilde{\alpha} \times 6\pi/k_0^3$, and omit the dependence on
$\omega$. We can now rewrite $\sigma_\mathrm{sc} = \sigma_0
|\tilde{\alpha}|^2$ and $\sigma_\mathrm{g}=\sigma_0
\left(|\tilde{\alpha}|^2-\mathrm{Im}(\tilde{\alpha})\right)$,
where $\sigma_0 = 6\pi/k_0^2$ is the resonant scattering
cross-section (for a $J=0 \rightarrow J=1$ transition), such that
the threshold condition, as expressed by equation
(\ref{eq.letokhov}), reduces to \cite{Froufe:2009}
\begin{equation}\label{eq.b0cr}
\rho \sigma_0 L_\mathrm{eff} =
\eta b_0  > \frac{2\pi}{\sqrt{3 |\tilde{\alpha}|^2 \, \left(
|\tilde{\alpha}|^2-\mathrm{Im}(\tilde{\alpha}) \right)}} \, ,
\end{equation}
where $b_0$ is the on-resonance optical thickness of the cloud.
This condition is valid as soon as the medium exhibits gain, i.e., $|\tilde{\alpha}|^2-\mathrm{Im}(\tilde{\alpha}) >
0$. Interestingly, the condition $\mathrm{Im}(\tilde{\alpha}) <
0$, corresponding to single-pass amplification ($T>1$), is not a
necessary condition.

The threshold condition is thus given by a critical on-resonance
optical thickness, which is an intrinsic parameter of the cloud,
expressed as a function of the complex atomic polarizability only,
which depends on the pumping parameters. Although the initial
condition of equation (\ref{eq.letokhov}) involves two
characteristic lengths, we emphasize here that this is really one
single independent parameter, as real and imaginary parts of the
atomic polarizability are related via Kramers-Kronig relations
\cite{Jackson}. This point is due to the originality of the system
that we are considering, in which the same atoms are used to
amplify and scatter light. This property can be fruitfully used to
experimentally determine the threshold, as only one single
measurement can provide enough information. A weak probe
transmission spectrum, which we can rewrite with our notations,
\begin{equation}\label{eq.trans}
T(\omega)= e^{-b_0\,
\mathrm{Im}\left[\tilde{\alpha}(\omega)\right]} \; ,
\end{equation}
allows indeed the full characterization of
$\tilde{\alpha}(\omega)$.

\subsection{Application to Mollow and Raman gain}\label{sect.threshold_evaluation}

We now apply the previous results to Mollow gain and to Raman
gain. For the first one we use the ab initio knowledge of the
polarizability of a strongly pumped two-level atoms. For the
second one, we use experimental transmission spectra $T(\omega)$
to extract the atomic polarizability via equation (\ref{eq.trans})
and Kramers-Kronig relations.

With Mollow gain, the polarizability is exactly and analytically
known \cite{Mollow:1972}, assuming a weak probe field, which is a
good hypothesis for calculating the threshold of the random laser:
\begin{equation}\label{eq.mollow}
\begin{split}
\tilde{\alpha}(\delta,\Delta,\Omega) = &-\frac{1}{2}\,
\frac{1+4\Delta^2}{1+4\Delta^2+2\Omega^2}  \, \\& \times
\frac{(\delta+i)(\delta-\Delta+i/2)-\Omega^2\delta/(2\Delta-i)}
{(\delta+i)(\delta-\Delta+i/2)(\delta+\Delta+i/2)-\Omega^2(\delta+i/2)}
\; ,
\end{split}
\end{equation}
where $\Delta$, $\delta$ and $\Omega$ are in unit of $\Gamma$.

\begin{figure}[t]
\centering
\includegraphics{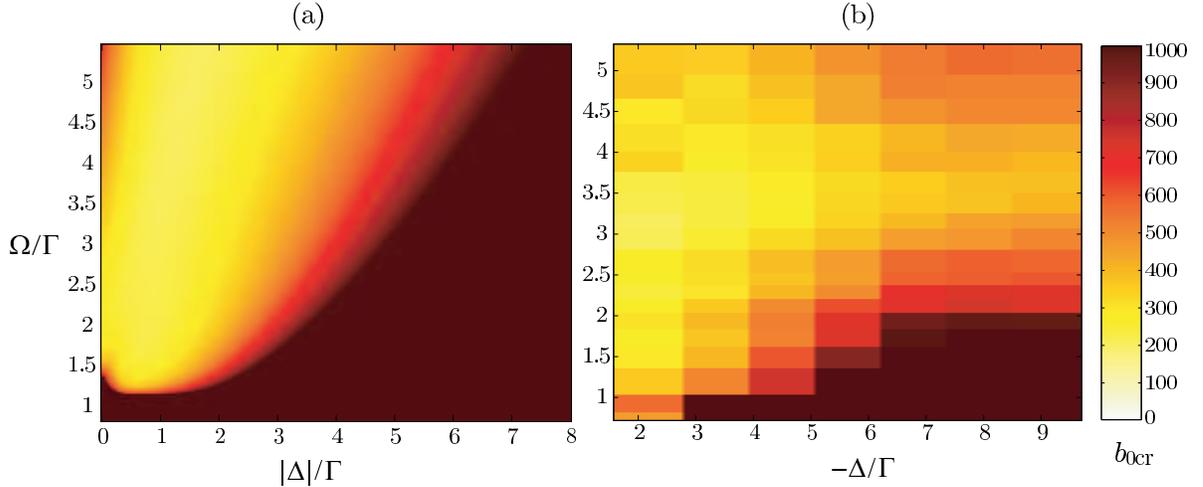}
\caption{Threshold of random lasing based on Mollow gain (left)
and Raman gain (right) for each pair of pumping parameters
$\Delta$ (detuning) and $\Omega$ (Rabi frequency). We use the
theoretical knowledge of the atomic polarizability
(\ref{eq.mollow}) for Mollow gain whereas the prediction for Raman
gain is based on experimental spectra (see text). In both cases,
we obtain a minimum optical thickness of about
200.}\label{fig.b0cr}
\end{figure}

For each pair of pumping parameters $\{\Delta, \Omega \}$, the use
of the polarizability (\ref{eq.mollow}) into the threshold
condition (\ref{eq.b0cr}) allows the calculation of the critical
on-resonance optical thickness $b_0$ as a function of the
pump-probe detuning $\delta$. Then, the minimum of $b_0$ and the
corresponding $\delta$ determine the optical thickness
$b_{0\mathrm{cr}}$ that the cloud must overcome to allow lasing,
and the frequency $\delta_\mathrm{RL}$ of the random laser at
threshold. The result is presented in figure \ref{fig.b0cr}(a) for
a spherical geometry ($\beta = 2$). The result for
$b_{0\mathrm{cr}}$ is independent of the sign of $\Delta$. The
minimum optical thickness that allows lasing is found to be
$b_{0\mathrm{cr}} \approx 200$ and is obtained for a large range
of parameters, approximately along the line $\Omega \approx 3
\Delta$. The optimum laser-pump detuning is near the gain line of
the transmission spectrum, i.e., $\delta_\mathrm{RL} \sim
\mathrm{sign}(\Delta) \sqrt{\Delta^2+\Omega^2}$, with however a
small shift compared to the maximum gain condition due to the
additional constraint of combined gain and scattering.

With Raman gain, we do not have a simple analytical expression of
the polarizability. Nevertheless, we can measure its imaginary
part by a transmission spectrum (equation (\ref{eq.trans})). A fit of the result, followed by
its transformation by Kramers-Kronig relations \cite{Jackson},
allows us to recover the full complex polarizability and to use
equation (\ref{eq.b0cr}) \cite{Guerin:2009}. Note that this method
is general and could be applied to any gain mechanism. However,
Raman gain is perfectly suited for this analysis as we can fit the
corresponding transmission profile by two inverted Lorentzian
lines (see the Raman structure on figure \ref{fig.Raman}(b), near
$\delta=0$), the transformation of which via Kramers-Kronig
relations is analytic and well-known (it gives a standard
dispersion profile). As the two Lorentzians are not well separated
(for most parameters, $\delta_\mathrm{R} < \gamma$), it leads for
the scattering cross-section ($\propto |\tilde{\alpha}|^2$) to a
bell-shaped curve centered near $\delta \sim 0$
\cite{Guerin:2009}. Note that besides creating gain, the Raman
transition \emph{adds} some scattering.

As for Mollow gain, we have varied the pump parameters $\{\Delta,
\Omega \}$ and for each pair, we have recorded a transmission
spectrum and deduced the corresponding random laser threshold. We
report the result on figure \ref{fig.b0cr}(b). The calculation
contains corrections due to the peculiar polarization used in the
experiment but deals only with the Raman resonance. It takes into
account neither the main atomic absorption line at $\omega_0$
(which adds elastic scattering), nor the losses due to inelastic
scattering. Doing so, the two corrections almost compensate each
other, and leads to $b_{0\mathrm{cr}} \sim 200$ as the correct
order of magnitude, obtained for $\Omega = \Delta \sim 3-4~\Gamma$
\cite{Guerin:2009}.

These results are very promising. Indeed, by using
well-established techniques to compress magneto-optical traps
(see, e.g., \cite{Ketterle:1993,DePue:2000}), achieving an optical
thickness of 200 should not be too difficult.

\subsection{Beyond the diffusive model}\label{sect.RTE}

The calculation of the lasing threshold based on Lethokov's
approach leads, for the optimum pumping parameters, to ratio
$L/\ell_\mathrm{sc} = b_{0\mathrm{cr}}|\tilde{\alpha}|^2 \sim 0.5$
for Mollow gain and $L/\ell_\mathrm{sc} \sim 2$ for Raman gain.
Because these values are not consistent {\it a priori} with the
domain of validity of the diffusion approximation \cite{Yoo:1990},
a more refined transport model is needed in order to check the
relevance of the approach. A model for random lasers with
incoherent feedback, going beyond the diffusion model, was
introduced by Noginov {\it et al.}~\cite{Noginov:2006}, based on a
phenomenological one-dimensional modelling of light transport. In
this work, we use a more advanced three-dimensional model based on
the Radiative Transfer Equation (RTE), that was introduced as a
tool to predict the threshold of ``classical'' random
lasers~\cite{Pierrat:2007}, and that we extended recently to
describe a cold-atom random laser with Mollow
gain~\cite{Froufe:2009}.

The RTE is a Boltzmann-type transport equation
\cite{Chandrasekhar}, that has a wider range of validity with
respect to the ratio $L/\ell_\mathrm{sc}$ than the diffusion
equation~\cite{Elaloufi:2004}. The basic quantity is the specific
intensity $I_{\omega}({\bf r},{\bf u},t)$, which describes the
number of photons at frequency $\omega$, at point ${\bf r}$,
propagating along direction ${\bf u}$ at time $t$. In order to
compare the diffusive and RTE predictions, we will consider a slab
geometry, with the $z$ axis normal to the slab surfaces. In the
case of plane-wave illumination at normal incidence, and for
isotropic scatterers, the specific intensity only depends on the
space variable $z$ and the angular variable $\mu=\cos\theta$, with
$\theta$ the angle between the propagation direction ${\bf u}$ and
the $z$-axis. In a system exhibiting gain and (isotropic)
scattering, the RTE reads:
\begin{equation}
 \frac{1}{c}   \, \frac{\partial I_\omega}{\partial t}(z,\mu,t)+ \mu \, \frac{\partial
 I_\omega}{\partial z}(z,\mu,t)= (\ell_\mathrm{g}^{-1} -
\ell_\mathrm{sc}^{-1})   \, I_\omega(z,\mu,t) +  (2 \, \ell_\mathrm{sc})^{-1}
\int_{-1}^{+1}  I_\omega(z,\mu^\prime,t) \, \mathrm{d}\mu^\prime
    \label{eq:ETR1}
\end{equation}
where $c$ is the energy velocity in the medium.

A feature of the RTE is that a modal expansion is available, whose
asymptotic behaviour at large length and time scales leads to the
modal expansion of the diffusion equation~\cite{Zweifel}.
Therefore, under the conditions of uniform (in space) and constant
(in time) gain, it is possible to build a modal theory of random
lasers with incoherent feedback based on the RTE, that generalizes
Lethokov's approach beyond the diffusive
regime~\cite{Pierrat:2007}. We focus on the slab geometry
($\beta=1$) since the modal expansion of the RTE is well known in
this case~\cite{Zweifel} (to our knowledge, no simple expansion is
available for a sphere in the RTE approach). The modal approach
consists in looking for solutions of the form $I(z,\mu,t)=
I_{\kappa,s}(\mu) \, \exp(i\kappa z) \, \exp(st)$, where
$I(z,\mu,t)$ is the specific intensity and the dependence on
$\omega$ has been omitted for simplicity. $\kappa$ can be chosen
as a real parameter, and $s$ can take complex values. For a given
real $\kappa$, $s(\kappa)$ and $I_{\kappa,s}$ form a set of
eigenvalues and eigenfunctions of the RTE. For isotropic
scattering, eigenvalues and eigenfunctions can be obtained
analytically~\cite{Zweifel}. In a passive medium, the eigenvalue
corresponding to the mode with the longest lifetime
reads\footnote{Note that a misprint occurred in the published
version of \cite{Froufe:2009} and that the correct expression is
given here.}:
\begin{equation}
s_0(\kappa)/c = \ell_\mathrm{g}^{-1} - \left[
\ell_\mathrm{sc}^{-1} - \kappa/\tan(\kappa\,\ell_\mathrm{sc}) \right]
 \  , \  \mathrm{for}  \ \kappa \, \ell_\mathrm{sc} <\frac{\pi}{2} \ .
\label{eq:s0RTE}
\end{equation}
In the presence of gain, the lasing threshold is reached when this
modes starts to display an exponential amplification in time.
Determining the threshold parameters amounts to calculating the
gain length and scattering mean free path generating an eigenvalue
$s_0(\kappa)>0$. Exactly at threshold, one has $s_0(\kappa)=0$.
For a slab of width $L$, the dominant mode corresponds to $\kappa
= \pi/L_\mathrm{eff} = \pi/(L +2\xi \ell_\mathrm{sc})$. In
practice, the determination of $\kappa$ is meaningful as long as
$\xi=0.71$ can be taken as a constant (independent on $L$).
Although not shown for brevity, we have verified with a full
numerical solution of the RTE that this is the case as soon as
$L>\ell_\mathrm{sc}$. This condition sets a limit of accuracy of
the modal approach.

It is well known that the diffusion approximation is
asymptotically reached from the RTE in the limit of long time and
large length scales. Indeed, in the limit $\kappa \ell_\mathrm{sc}
\ll 1$, a first-order expansion of equation~(\ref{eq:s0RTE})
yields
\begin{equation}
s_0^\mathrm{(DA)}(\kappa)/c =\ell_\mathrm{g}^{-1} -\kappa^2
\ell_\mathrm{sc}/3,
\label{eq:s0DA}
\end{equation}
which is the dispersion relation of the modes of the diffusion
equation (DA)~\cite{Zweifel,Pierrat:2006}. The associated
threshold condition $s_0^\mathrm{(DA)}(\kappa=\pi/L_\mathrm{eff})
= 0$ corresponds exactly to the result obtained with Letokhov's
approach, with $\beta=1$. A careful look at equations
(\ref{eq:s0RTE}) and (\ref{eq:s0DA}) shows that the predictions of
the RTE and diffusion approximation should not be substantially
different. Firstly, we note that the gain contribution to
$s_0(\kappa)$ is the same in both models (the first term is the
same in equations (\ref{eq:s0RTE}) and (\ref{eq:s0DA})). Secondly,
the scattering contribution (the second term in the equations) is
larger in the RTE model, but by a factor that remains smaller than
$1.13$ (when $L \sim \ell_\mathrm{sc}$). This means that the
correction introduced by the RTE remains relatively small, at
least for the slab geometry. Numerically, it corresponds to an
increase of $\eta \, b_{0\mathrm{cr}}$ of at most a few
percents~\cite{Froufe:2009}. We can conclude that the model based
on the diffusion approximation gives accurate results, even in the
regime $L \sim \ell_\mathrm{sc}$. Also note that the accuracy
should be better with Raman gain, for which the optimum conditions
are reached for $L/\ell_\mathrm{sc}\sim 2$ (instead of
$L/\ell_\mathrm{sc}\sim 0.5$ for Mollow gain). Finally, we stress
that our prediction shows that in a cold-atom system, random
lasing could be achieved even in a regime of low scattering. This
is a feature of a system exhibiting a high level of gain, in which
the threshold can be reached even with a low-quality feedback
(i.e., a cavity with a poor quality factor).

\subsection{Emitted intensity above threshold}\label{sect.power}

In this section we focus on the characterization of the Mollow
random laser above threshold. The threshold of a random laser with
Mollow gain can be predicted using the lasing condition
(\ref{eq.b0cr}) together with the polarizability given by equation
(\ref{eq.mollow}). This polarizability is obtained in the weak
probe limit \cite{Mollow:1972}. When the lasing threshold is
surpassed, Letokhov's theory leads  to an exponential growth of
the laser intensity versus time, and hence steady state cannot be
reached. In order to avoid this unphysical effect, as already
pointed in \cite{Letokhov:1968}, saturation effects must be
included in the description of the atomic polarizability at both
pump and probe frequencies.

To do so, optical Bloch equations in the strong probe regime can
be numerically solved in order to obtain the atomic
polarizability. In this case the atomic polarizability depends on
the lasing intensity in the medium $I_\mathrm{RL}^\mathrm{(in)}$
through its associated Rabi frequency $|\Omega_\mathrm{RL}|^ 2
\propto I_\mathrm{RL}^\mathrm{(in)}$. Hence,
$\tilde{\alpha}=\tilde{\alpha}(\delta, \Delta,
\Omega,\Omega_\mathrm{RL})$ (all frequencies will be in unit of
$\Gamma$ in the following equations). The steady-state value of
the random laser Rabi frequency is considered as the value at
which losses exactly compensate gain. In this situation, the
output power of the random laser $P_\mathrm{RL}^\mathrm{(out)}$
equals the generated power in the lasing medium, i.e.,
$P_\mathrm{RL}^\mathrm{(out)} \propto \sigma_\mathrm{g}
|\Omega_\mathrm{RL}|^2$, with $\sigma_\mathrm{g} = \sigma_0
\left(|\tilde{\alpha}|^2-\mathrm{Im}(\tilde{\alpha}) \right)$ the
gain cross-section. On the other hand, the pump-induced
fluorescence power is $P_\mathrm{Fluo} \propto \sigma_0
|\Omega|^2/(1+4\Delta^2+2|\Omega|^2)$.

\begin{figure}[t]
\centering
\includegraphics{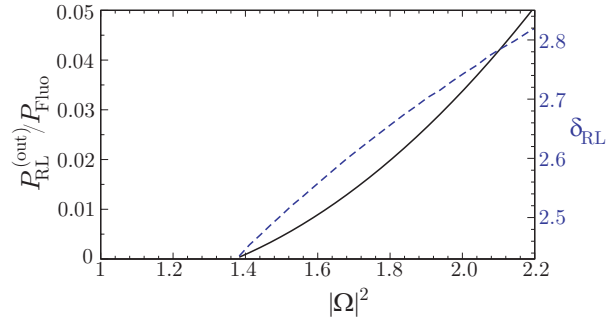}
\caption{Continuous line: Emitted random laser power normalized to
the pump fluorescence power, as a function of the pump intensity.
Dashed line: Normalized laser detuning $\delta_\mathrm{RL}$. The
random medium is a spherical cloud of two level atoms with an
on-resonance optical-thickness $b_{0}=600$.}\label{fig:power}
\end{figure}

Hence, in this regime the ratio of the lasing power to the
fluorescence power induced by the pump can be estimated. This
ratio is an important step towards the performance
characterization of the random laser, as it allows to quantify the
amount of laser signal that can be extracted from the fluorescence
background. From the previous results we get

\begin{equation}\label{eq.intensite}
\frac{P_\mathrm{RL}^\mathrm{(out)}}{P_\mathrm{Fluo}}=
\frac{|\Omega_\mathrm{RL}|^2}{|\Omega|^2}
\left(|\tilde{\alpha}|^{2}-\textrm{Im}(\tilde{\alpha})\right)
\left(1+4\Delta^2+2|\Omega|^2\right) \; .
\end{equation}

As can be seen in figure \ref{fig:power}, the ratio of lasing
power to fluorescence background is on the order of 5\% for a pump
detuning $\Delta=1$, leading  to a measurable signal. One
interesting characteristic of this random laser concerns the laser
frequency at its maximum emission power. As shown in figure
\ref{fig:power}, this maximum emission frequency shifts as the
pump intensity increases. This effect is due to the shift in the
Mollow gain as the pumping power changes.

\subsection{Limitations of the model}\label{sect.limit}

Of course these crude estimations neglect different effects. We
assume homogeneous pump intensity across the whole system
\cite{Noginov:2006}. We are considering one pump frequency and one
lasing frequency, neglecting thus effects resulting from mode
competition \cite{Tureci:2008} and inelastic scattering of laser
light \cite{Khaykovich:1999} among others. Despite neglecting
those effects, these estimations seem to be reasonable at least in
recovering the orders of magnitude of the lasing threshold and
power.

The inhomogeneities in the atomic cloud density, in the pump
intensity and in the gain distributions could be incorporated in
the RTE model, at the cost of a full numerical treatment of
coupled equations for the pump beam and emitted light intensities.
Such an approach has been developed previously in the case of a
``classical'' random laser, i.e., with gain and scattering as
separate entities~\cite{Pierrat:2007}. Another improvement would
be to deal with a spherical geometry, that is closer to the real
conditions. Finally, including the Raman gain mechanism into the
RTE approach, using experimental data as input parameters, would
be of great interest in view of experiments since the threshold
conditions seem easier to reach than with Mollow gain, in view of
our first estimates.

\section{Outlook and conclusion}\label{sect.conclusion}

We have presented in this paper our recent investigations on the
issue of achieving a diffusive random laser in a cloud of cold
atoms. We have especially studied different gain mechanisms in a
situation of standard laser, and we have quantified the random
laser threshold for two of them. These evaluations show that
random laser in cold atoms \emph{is possible}. This is undoubtly
our main result. Moreover, we point out that the peculiarity of
our system would lead to a random laser with a low amount of
scattering, that is, low feedback. This regime is similar to that
encountered in certain semiconductor lasers with a very poor
cavity, and is different from the working regime of random lasers
realized to date.

Moreover, from our experience, we can now draw some preliminary
conclusions about the comparison of the different gain mechanisms.
Application of Letokhov's criteria leads to similar critical
optical thickness for Mollow gain and Raman gain. Nevertheless,
the optimum for Mollow gain is obtained outside the range of
validity of our transport model whereas the amount of scattering
is larger with Raman gain. Moreover, polarization is better taken
into account in our prediction for Raman gain. Finally, the
optimum pumping parameters are such that pump penetration through
the cloud will be higher with Raman gain, since detuning and power
are larger. For all these reasons, Raman gain seems more
appropriate to achieve random lasing.

However, achieving random lasing is not enough, as we need also to
\emph{detect} the random laser emission. In our system, gain is
almost at the same wavelength than the pump (and thus its
fluorescence), and this makes the detection challenging. In this
respect, Raman gain is worse than Mollow gain, as the very small
detuning $\delta_\mathrm{R}$ between the pump and the gain
frequency prevent the use of a Fabry-Perot interferometer to
distinguish between the laser emission and the pump-induced
fluorescence. This could be done more easily with Mollow gain
(see, e.g., \cite{Wu:1975}). Moreover, from our preliminary
evaluations of the intensity of the random laser emission with
Mollow gain (section \ref{sect.power}), we know that the random
laser emission should not be too small to be detected.

For Raman gain we can imagine other ways to extract the optical
spectrum of the emitted light, by looking at the intensity
correlations of the fluorescence, measured by a heterodyne
technique \cite{Westbrook:1990} or a homodyne technique
\cite{Jurczak:1995} or a correlator \cite{Bali:1996}. Nevertheless
these techniques do not enable to filter the random laser light,
contrary to the use of a Fabry-Perot cavity. Filtering the random
laser light could be very useful to study the random laser
properties, for example the intensity temporal correlations. Note
that this kind of experiments could be interesting also below the
random laser threshold \cite{Yamilov:2005,Fedorov:2009}.

The third gain mechanism that we have studied is four-wave mixing,
which has lead to a very efficient cavity-laser. Nevertheless, the
configuration we have used so far, that is \emph{degenerate} FWM,
is not appropriate at all for the random laser problem, as the
pump fluorescence and the random laser emission would be exactly
at the same frequency. However, another configuration of
parametric gain is possible, that is non-degenerate FWM, for which
two different frequencies $\omega_\mathrm{F}$ and
$\omega_\mathrm{B}$ are used for the two pump fields, and
amplification occurs at frequency $\omega =
\left(\omega_\mathrm{F}+\omega_\mathrm{F}\right)/2$. Then, it
should be possible to choose a large enough detuning to facilitate
the detection, and to adjust the gain frequency close to the
atomic frequency $\omega_0$ to enhance scattering. Preliminary
numerical simulations indicate that this configuration is
promising \cite{Nicolas:rapport}, and its experimental test
constitutes our next work.

Finally, we may also use some additional degrees of freedom that atoms
offer. For example, cold atomic clouds have been shown to exhibit
large Faraday rotations in presence of a magnetic field
\cite{Labeyrie:2001}. This Faraday effect may change the threshold
of the random laser \cite{Pinheiro:2008}. Moreover, magnetic
fields can change also the atomic radiation pattern
\cite{Labeyrie:2002} and the coherence length of the light
transport inside the sample \cite{Sigwarth:2004}. These effects
could thus be used to engineer the multiple scattering properties
of atoms and therefore to achieve a better control on the random
laser based on cold atoms.

\ack This work is supported by ANR CAROL (project
ANR-06-BLAN-0096). L.S.F. acknowledges the financial support of
Spanish ministry of science and innovation through its Juan de la
Cierva program, D.B. acknowledges support from INTERCAN, N.M. and
F.M. are funded by DGA.

\appendix

\section{Laser dynamics equations} \label{appendix}

\setcounter{section}{1}

We give here the expressions of the functions $\Phi$, $\Psi$ and
$\Theta$ that describe the coupling terms in the laser dynamics
coupled equations (\ref{eq.pop}--\ref{eq.champ})
\cite{Zakrzewski:1992_1,Zakrzewski:1991}:
\begin{eqnarray}
& \Phi(\sigma_j, a_m)  =
2i\lambda_{1}\sum_{m=1}^{M}\left(\exp[-i\phi_{j}]\sigma_{j}a_{m}^{*}-\exp[i\phi_{j}]\sigma_{j}^{*}a_{m}\right)
 \label{eq.Phi}\\
 & \quad +
 2i\lambda_{2}\sum_{m=1}^{M}\left(\exp[-i\phi_{j}]\sigma_{j}^{*}a_{m}^{*}-\exp[i\phi_{j}]\sigma_{j}a_{m}\right)+2\gamma_{3}\left(\sigma_{j}+\sigma_{j}^{*}\right)\;
 ,
\nonumber
\\
& \Psi(\sigma_j, \sigma_{3j}, a_m)  =
-i\sum_{m=1}^{M}\left(\exp[i\phi_{j}]\lambda_{1}\sigma_{3j}a_{m}+\exp[-i\phi_{j}]\lambda_{2}\sigma_{3j}a_{m}^{*}\right)
\label{Psi} \\
 & \quad -
 2i\lambda_{0}\sum_{m=1}^{M}\left(\exp[-i\phi_{j}]\sigma_{j}a_{m}^{*}+\exp[i\phi_{j}]\sigma_{j}a_{m}\right)+\gamma_{0}+\gamma_{3}\sigma_{3j}-\gamma_{4}\sigma_{j}^{*}\;
 ,
\nonumber
\\
& \Theta(\sigma_{3j}, \sigma_j)  =
-i\sum_{j=1}^{N}\exp[-i\phi_{j}]\left(\lambda_{1}\sigma_{j}-\lambda_{2}\sigma_{j}^{*}+\lambda_{0}\sigma_{3j}\right)\; , &
\label{eq.Theta}
\end{eqnarray}
where $\gamma_{0}=(\Gamma/2)\sin2\theta$,
$\gamma_{3}=(\Gamma/4)\sin2\theta\cos2\theta$,
$\gamma_{4}=(\Gamma/4)\sin^{2}2\theta$,
$\lambda_{0}=(g/4)\sin2\theta$,
$\lambda_{1}=(g/4)\left(1+\cos2\theta\right)$ and
$\lambda_{2}=(g/4)\left(1-\cos2\theta\right)$.

The procedure to numerically solve the equations
(\ref{eq.pop}--\ref{eq.champ}) follows the method presented in
\cite{Zakrzewski:1992_1}. We introduce a family of $n$-photon
polarizations $S_n=\sum_{j}\exp[-i n \phi_{j}]\sigma_{j}$ and
$S_{3n}=\sum_{j}\exp[-i n \phi_{j}]\sigma_{3j}$. Equations
(\ref{eq.pop},\ref{eq.coh}) cast then into an infinite set of
hierarchical equations for $S_n$ and $S_{3n}$. In this paper, we
truncate this set of equations at $|n| = 5$, which, in the case of
$M = 1$, leaves us with a total of 35 equations and corresponds to
the generalized effective theory GET35, if we use terminology of
\cite{Zakrzewski:1992_1}. The hierarchical equations for
macroscopic polarizations $S_n$ and $S_{3n}$, supplemented with
equations for amplitudes of modes $a_m$ of the electromagnetic
field, are then solved numerically.

\section*{References}


\providecommand{\newblock}{}

\end{document}